\documentclass[letterpaper,10pt,conference]{ieeeconf}
\IEEEoverridecommandlockouts
\usepackage{amsmath}
\usepackage{graphicx}

\usepackage{amsfonts}
\usepackage{color}
\usepackage{epstopdf} % for postscript graphics files
\newcommand*{\red}{\textcolor{black}}
\newcommand{\smallsub}[1]{\! {\scriptscriptstyle \mathcal{#1}} }

\newtheorem{proposition}{Proposition}

\newtheorem{definition}{Definition}
\newtheorem{remark}{Remark}
\newtheorem{lemma}{Lemma}
\newtheorem{assumption}{Assumption}

\newtheorem{example}{Example}

\newtheorem{theorem}{Theorem}
\newtheorem{property}{Property}

\newcommand{\Eb}{\bar{\mathbb{E}}}
\newcommand{\M}{\mathcal{M}}
\newcommand{\E}{\mathbb{E}}

\newcommand{\mY}{\mathcal{Y}}
\newcommand{\mD}{\mathcal{D}}
\newcommand{\mZ}{\mathcal{Z}}
\newcommand{\mQ}{\mathcal{Q}}
\newcommand{\mA}{\mathcal{A}}
\newcommand{\mB}{\mathcal{B}}
\newcommand{\mN}{\mathcal{N}}
\newcommand{\mL}{\mathcal{L}}
\newcommand{\smX}{\smallsub{X}}
\newcommand{\smZ}{\smallsub{Z}}

\newcommand{\smA}{\smallsub{A}}
\newcommand{\smB}{\smallsub{B}}

\newcommand{\smD}{\smallsub{D}}
\newcommand{\smY}{\smallsub{Y}}
\newcommand{\smQ}{\smallsub{Q}}
\newcommand{\beq}{\begin{equation}}
\newcommand{\eeq}{\end{equation}}
\newcommand{\beqs}{\begin{equation*}}
\newcommand{\eeqs}{\end{equation*}}
\newcommand{\beqr}{\begin{eqnarray}}
\newcommand{\eeqr}{\end{eqnarray}}
\newcommand{\beqrs}{\begin{eqnarray*}}
\newcommand{\eeqrs}{\end{eqnarray*}}
\newcommand*{\blue}{\textcolor{black}}

\title{\LARGE \bf Local module identification in dynamic networks \\ with correlated noise: the full input case}

\author{Paul M.J. Van den Hof, Karthik R. Ramaswamy, Arne G. Dankers and Giulio Bottegal % <-this % stops a space
\thanks{Paul Van den Hof, Karthik Ramaswamy and Giulio Bottegal are with the Department of Electrical
Engineering, Eindhoven University of Technology, Eindhoven, The
Netherlands {\tt\small \{p.m.j.vandenhof, k.r.ramaswamy, g.bottegal\}@tue.nl}}
\thanks{Arne Dankers is with the Electrical and Computer Engineering Dept. at the University of Calgary, Canada, {\tt\small adankers@hifieng.com}}
\thanks{This work has received funding from the European Research Council (ERC), Advanced Research Grant SYSDYNET, under the European Union's Horizon 2020 research and innovation programme (grant agreement No 694504).}
}

\begin{document}

\maketitle
\thispagestyle{empty}
\pagestyle{empty}

\begin{abstract}
The identification of local modules in dynamic networks with known topology has recently been addressed by formulating conditions for arriving at consistent estimates of the module dynamics, typically under the assumption of having disturbances that are uncorrelated over the different nodes. The conditions typically reflect the selection of a set of node signals that are taken as predictor inputs in a MISO identification setup. In this paper an extension is made to arrive at an identification setup for the situation that process noises on the different node signals can be correlated with each other. In this situation the local module may need to be embedded in a MIMO identification setup for arriving at a consistent estimate with maximum likelihood properties. This requires the proper treatment of confounding variables. The result is an algorithm that, based on the given network topology and disturbance correlation structure, selects an appropriate set of node signals as predictor inputs and outputs in a MISO or MIMO identification setup. As a first step in the analysis, we restrict attention to the (slightly conservative) situation where the selected output node signals are predicted based on all of their in-neighbor node signals in the network.
\end{abstract}

\section{INTRODUCTION} \label{sec-intro}
In recent years increasing attention has been given to the development of new tools for the identification of large-scale interconnected systems, also known as dynamic networks. These networks are typically thought of as a set of measurable signals (the node signals) interconnected through linear dynamic systems (the modules), possibly driven by external excitations (the reference signals). Among the literature on this topic, we can distinguish three main categories of research. The first one focuses on identifying the topology of the dynamic network \cite{MaterassiSalapaka2012}, \cite{Sanandaji2011}, \cite{Materassi10}, \cite{Chiuso&Pillonetto_Autom:12}. The second category concerns  identification of the full network dynamics \cite{Weerts&etal_Autom:18_reducedrank}, \cite{Weerts&etal_CDC:16}, \cite{Weerts&etal_Autom:18_identifiability}, \cite{Bazanella&etal_CDC:17} while the third one deals with identification of a specific component (module) of the network, assuming that the network topology is known (the so called local module identification, see \cite{VandenHof&etal_Autom:13}, \cite{Ramaswamy&etal_CDC:18}, \cite{Everitt&Bottegal&Hjalmarsson_Autom:18}, \cite{Gevers&etal_SYSID:18}, \cite{Dankers&etal_Autom:15}).

In this paper we will further expand the work on the local module identification problem. In \cite{VandenHof&etal_Autom:13}, the classical \emph{direct-method} \cite{Ljung:99} for closed-loop identification has been generalized to a dynamic network framework using a MISO identification setup. Consistent estimates of the target module can be obtained when the network topology is known and all the node signals in the MISO identification setup are measured. The work has been extended in \red{\cite{Materassi&etal_CDC:15,Dankers&etal_TAC:16}} towards the situation where some node signals might be non-measurable, leading to an additional predictor input selection problem. A similar setup has also been studied in \cite{Ramaswamy&etal_CDC:18}, where an approach has been presented based on empirical Bayesian methods to reduce the variance of the target module estimates. In \cite{Dankers&etal_Autom:15} and \cite{Everitt&Bottegal&Hjalmarsson_Autom:18}, dynamic networks having node measurements corrupted by sensor noise have been studied, and informative experiments for consistent local module estimates have been addressed in \cite{Gevers&etal_SYSID:18}.

%A generalization of the \emph{two-stage method} \cite{VandenHof&Schrama_Autom:93} to a dynamic network framework is addressed in \cite{VandenHof&etal_Autom:13} and \cite{Dankers&etal_TAC:16} where a MISO identification setup is used to obtain consistent estimate of the target module under the situation of having disturbance signals at different nodes to be correlated. Even though these methods can handle correlated noise and provide consistent estimates, the obtained estimates have non-optimal variance.

A standing assumption in the aforementioned works \cite{VandenHof&etal_Autom:13}, \cite{Ramaswamy&etal_CDC:18}, \cite{Gevers&etal_SYSID:18}, \cite{Dankers&etal_TAC:16} is that the process noises entering the nodes of the dynamic network are uncorrelated with each other. This assumption facilitates the analysis and the development of methods for local module identification, reaching
%because only local node measurements are required to be set as either predicted outputs or predictor inputs
%and required to build an identification setup that achieves
\emph{consistent} module estimates using the direct method. However, when process noises are correlated over the nodes, the consistency results for the considered MISO direct method collapse. In this situation it is seems necessary to consider also the \emph{noise topology or disturbance correlation structure}, when selecting an appropriate identification setup.  Even though the two-stage methods in \cite{Dankers&etal_Autom:15} and \cite{Everitt&Bottegal&Hjalmarsson_Autom:18} can handle the situation of correlated noise and deliver consistent estimates, the obtained estimates will not have minimum variance.

In this paper we precisely consider the situation of having dynamic networks with disturbance signals on different nodes that possibly are correlated, while our target moves from consistency only, to also minimum variance (or Maximum Likelihood (ML)) properties of the obtained estimates. While one could use techniques for full network identification (e.g., \cite{Weerts&etal_Autom:18_reducedrank}), our aim is to develop a method that uses only local information. In this way, we avoid (i) the need to collect node measurements that are ``far away'' from the target module, and (ii) the need to identify unnecessary modules that would come with the price of higher variance in the estimates. \red{We will assume that the topology of network is known, as well as the correlation structure of the noise disturbances.}

%, while our target moves from consistency only to also minimum variance (or ML) properties of the obtained estimates. For identification of a full network, it has been shown in \cite{Weerts&etal_Autom:18_reducedrank} that this requires a multi-input multi-output (MIMO) identification setup. A generalization of the direct method called \emph{joint-direct method} is introduced in \cite{Weerts&etal_Autom:18_reducedrank} where all nodes in the dynamic network are used as both predictor inputs and predicted outputs. This has been confirmed in \cite{VandenHof&etal_CDC:17}, where it has been shown that for a simple two-node network with correlated disturbances, even for consistent identification of a local module a multi-output identification setup is necessary. But in both \cite{Weerts&etal_Autom:18_reducedrank} and \cite{VandenHof&etal_CDC:17}, since we include all node signals in the dynamic network as predictor inputs and predicted outputs, we need to perform a full network identification even if we would want to identify a particular module.

Using the reasoning first introduced \cite{VandenHof&etal_CDC:17}, we build a constructive procedure that, choosing a limited number of predictor inputs and predicted outputs, builds an identification setup that guarantees maximum likelihood (ML) properties (and thus asymptotic minimum variance) when applying a direct prediction error identification method. In this situation we have to deal with so-called \emph{confounding variables} (see e.g. \cite{VandenHof&etal_CDC:17}, \blue{\cite{Dankers&etal_IFAC:17}}), that is, unmeasured variables that directly or indirectly influence both the predicted output and the predictor inputs, and lead to lack of consistency. A direct influence, caused by correlated process noise, can be treated by adding predicted outputs to our identification setting, while an indirect influence, caused by unmeasured nodes, can be resolved by adding predictor inputs. In this paper, we restrict our attention to the situation where all the nodes that are in-neighbors of predicted outputs are measured, which we refer to as the \emph{full input case}.

This paper is organized as follows. In section II, the dynamic network setup is defined. Section III provides a summary of available results from the existing literature of local module identification related to the context of this paper. Next, important concepts and notations used in this paper are defined in Section IV. Section V provides an algorithm for selecting the predictor inputs and predicted outputs while the MIMO identification setup and predictor model are provided in the next section. Section VII presents the main results of this paper followed by two illustrative examples of the introduced method in the subsequent section. Conclusions are discussed in section IX.

\section{Network \red{and identification} setup}

Following the basic setup of \cite{VandenHof&etal_Autom:13}, a dynamic network is built up out of $L$ scalar \emph{internal variables} or \emph{nodes} $w_j$, $j
= 1, \ldots, L$, and $K$ \emph{external variables} $r_k$, $k=1,\ldots K$.
Each internal variable is described as:
\vspace{-0.15cm}
\begin{align}
w_j(t) = \sum_{\stackrel{l=1}{l\neq j}}^L
G_{jl}(q)w_l(t) + r_j(t) + v_j(t)
\label{eq:netw_def}
\end{align}
\vspace{-0.3cm}

\noindent where $q^{-1}$ is the delay operator, i.e. $q^{-1}w_j(t) = w_j(t-1)$;
\begin{itemize}
	\item $G_{jl}$ is a proper rational transfer function matrix, and the single transfer functions $G_{jl}$ are referred to as {\it modules}.
	\item $r_j$ are \emph{external variables} that can directly be manipulated by the user and that may or may not be present; if $r_j$ is not present it is replaced by $r_j=0$.
	\item $v_j$ is \emph{process noise}, where the vector process $v=[v_1 \cdots v_L]^T$ is modelled as a stationary stochastic process with rational spectral density \red{$\Phi_v(\omega)$}, such that there exists a white noise process $e:= [e_1 \cdots e_L]^T$, with covariance matrix $\Lambda>0$ such that
$ v(t) = H(q)e(t)$,
where $H$ is square, stable, monic and minimum-phase. \red{The situation of correlated noise, as considered in this paper, refers to the situation that $\Phi_v(\omega)$ and $H$ are non-diagonal, while we assume that we know a priori which entries of $\Phi_v$ are nonzero.}
\end{itemize}
\red{We will assume that the standard regularity conditions on the data are satisfied that are required for convergence results of prediction error identification method\footnote{\red{See \cite{Ljung:99} page 249. This includes the property that $e(t)$ has bounded moments of order higher than $4$.}}.}
In line with the situations considered in \cite{Weerts&etal_Autom:18_identifiability}, we will assume that either all modules in $G$ are strictly proper, or that $\Lambda$ is restricted to be diagonal.

When combining the $L$ node signals we arrive at the full network expression
%\vspace{-0.2cm}
\begin{align*}
\begin{bmatrix}  \! w_1 \!  \\[1pt] \! w_2 \!  \\[1pt]  \! \vdots \! \\[1pt] \! w_L \!  \end{bmatrix} \!\!\! = \!\!\!
\begin{bmatrix}
0 &\! G_{12} \!& \! \cdots \! &\!\! G_{1L} \!\\
\! G_{21} \!& 0 & \! \ddots \! &\!\!  \vdots \!\\
\vdots &\! \ddots \!& \! \ddots \! &\!\! G_{L-1 \ L} \!\\
\! G_{L1} \!&\! \cdots \!& \!\! G_{L \ L-1} \!\! &\!\! 0
\end{bmatrix} \!\!\!\!
\begin{bmatrix} \! w_1 \!\\[1pt]  \! w_2 \!\\[1pt] \! \vdots \!\\[1pt] \! w_L \! \end{bmatrix} \!\!\!
+ \!\!
\begin{bmatrix} \! r_1 \!\\[1pt] \! r_2 \!\\[1pt] \! \vdots \!\\[1pt]  \! r_{K} \!\end{bmatrix}
\!\!\!+\!\!
H  \!\! \begin{bmatrix}\! e_1 \!\\[1pt] \! e_2 \!\\[1pt] \! \vdots \!\\[1pt] \! e_L\!\end{bmatrix} \!\!\!
\end{align*}
which results in the matrix equation:
\begin{align} \label{eq.dgsMatrix}
w = G w + r + H e.
\end{align}

\noindent The identification problem to be considered is the problem of identifying one particular module $G_{ji}(q)$ on the basis of measured variables $w$, and possibly $r$. In the current approaches to this problem, attention has been given to the selection of predictor input variables, when the target is to identify module $G_{ji}^0$ consistently.

Let us define $\mathcal{N}_j$ as the set of node indices $k$ such that $G_{jk} \neq 0$, i.e. the node signals in $\mathcal{N}_j$  are the in-neighbors of the node signal $w_j$.
Let $\mathcal{D}_j$ denote the set of indices of the internal variables that
are chosen as predictor inputs.
%\textcolor{green}{, i.e. the internal variable $w_k$ is a predictor input if and only if $k \in
%\mathcal{D}_j$}.
Let
$\mathcal{Z}_j$ denote the set of indices not in $\{j\} \cup \mathcal{D}_j$, i.e. $\mathcal{Z}_j = \{1, \ldots, L\} \setminus \{ \{j\} \cup \mathcal{D}_j \}$.  Let $w_{\smallsub{D}}$ denote the vector $[w_{k_1} \ \cdots \
w_{k_n} ]^T$, where $\{k_1,\ldots,k_n \} = \mathcal{D}_j$. Let
$r_{\smallsub{D}}$ denote the vector $[r_{k_1} \ \cdots \ r_{k_n}]^T$, where $\{k_1,\ldots,k_n\} = \mathcal{D}_j$, and where the
$\ell$th entry is zero if $r_{\ell}$ is not present in the
network. The vectors
$w_{\smallsub{Z}}$, $v_{\smallsub{D}}$, $v_{\smallsub{Z}}$ and
$r_{\smallsub{Z}}$ are defined analogously. The ordering of the
  elements of $w_{\smallsub{D}}$, $v_{\smallsub{D}}$, and
  $r_{\smallsub{D}}$ is not important, as long as it is the same for
  all vectors. The transfer function
  matrix between $w_{\smallsub{D}}$ and $w_j$ is denoted
$G_{j\smallsub{D}}^0$. The other transfer function
matrices are defined analogously.

\red{To illustrate the notation, consider the network sketched in Figure \ref{fig1}, and let module $G_{21}^0$ be the target module for identification.
\vspace{-0.41cm}
\begin{figure}[htb]
\centerline{\includegraphics[scale=0.4]{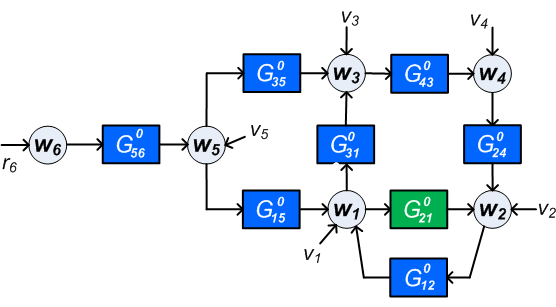}}
\vspace{-0.35cm}
	\caption{Example network}
	\label{fig1}
\end{figure}
Then $j=2$, $i=1$; $\mathcal{N}_j = \{1,4\}$. If we choose the set of predictor inputs as $\mathcal{D}_j = \mathcal{N}_j$, then the set of remaining (nonmeasured) signals, becomes $\mathcal{Z}_j = \{ 3,5,6\}$.}

By this notation, the network equations (\ref{eq.dgsMatrix}) is rewritten as:
\begin{align} \label{eq.dgsPartitioned}
\begin{bmatrix} w_j \\ w_{\smallsub{D}} \\
w_{\smallsub{Z}} \end{bmatrix} =
\begin{bmatrix} 0 & G_{j \smallsub{D}}^0 & G_{j \smallsub{Z}}^0 \\
G_{\smallsub{D} j}^0 & G_{\smallsub{D} \smallsub{D}}^0 & G_{\smallsub{D}
\smallsub{Z}}^0 \\ G_{\smallsub{Z} j}^0 &
G_{\smallsub{Z} \smallsub{D}}^0 & G_{\smallsub{Z}
\smallsub{Z}}^0 \end{bmatrix}
\begin{bmatrix} w_j \\ w_{\smallsub{D}} \\
w_{\smallsub{Z}} \end{bmatrix} +
\begin{bmatrix} v_j \\ v_{\smallsub{D}} \\
v_{\smallsub{Z}} \end{bmatrix} +
\begin{bmatrix} r_j \\ r_{\smallsub{D}} \\
r_{\smallsub{Z}} \end{bmatrix},
\end{align}
where $G_{\smallsub{D} \smallsub{D}}^0$ and $G_{\smallsub{Z}
\smallsub{Z}}^0$ have zeros on the diagonal.

Identification of module $G_{ji}^0$ can now be done by selecting $\mathcal{D}_j$ such that $i \in \mathcal{D}_j$, and subsequently estimating a multiple-input single output model \red{for the transfer functions in $G_{j\smallsub{D}}$}. This can be done by considering the \red{one-step-ahead predictor\footnote{$\Eb$ refers to $\lim_{N\rightarrow\infty} \frac{1}{N} \sum_{t=1}^N \E$, and $w_j^{\ell}$ and $w_{\mathcal{D}_j}^{\ell}$ refer to signal samples $w_j(\tau)$ and $w_k(\tau)$, $k\in \mathcal{D}_j$, respectively, for all $\tau \leq \ell$.}
\[ \hat w_j(t|t-1):= \Eb \{w_j(t)\ |\ w_j^{t-1},w_{\mathcal{D}_j}^t\}  \]
and the resulting} prediction error (\cite{Ljung:99}):
\vspace{-0.15cm}
\begin{align} \label{eq.predictionError}
\varepsilon_j(t,&\theta) =
w_j(t) - \hat{w}_j(t|t-1,\theta) \nonumber \\
&\hspace{-0.5cm} =  H_j(\theta)^{-1} \Big ( w_j - \!\! \sum_{k \in \mathcal{D}_j}
G_{jk}(\theta) w_k -  r_j \Big)
\end{align}
\vspace{-0.4cm}

\noindent where arguments $q$ and $t$ have been dropped for notational
clarity. The parameterized transfer functions $G_{jk}(\theta)$, $k \in
\mathcal{D}_j$ and
$H_j(\theta)$ are estimated by
minimizing the sum of squared (prediction) errors:
$V_j(\theta) = \frac{1}{N}
\sum_{t=0}^{N-1} \varepsilon_j^2(t,\theta),
$
%\vspace{-0.25cm}
%\begin{align} \label{eq.sse} V_j(\theta) = \frac{1}{N}
%\sum_{t=0}^{N-1} \varepsilon_j^2(t,\theta),
%\end{align}
%\vspace{-0.35cm}
where $N$ is the length of the data set. We refer to this identification method as the {\it direct method}, \cite{VandenHof&etal_Autom:13}. Let $\hat{\theta}_N$ denote the minimizing argument of $V_j(\theta)$.
%(\ref{eq.sse}).

\section{Available results}

The following results are available from previous work:
\begin{itemize}
\item When $\mathcal{D}_j$ is chosen equal to $\mathcal{N}_j$ and noise $v_j$ is uncorrelated to all $v_k$, $\red{k \neq j}$, then $G_{ji}^0$ can be consistently estimated in a MISO setup, provided that there is enough excitation in the predictor input signals, see \cite{VandenHof&etal_Autom:13}.
\item When $\mathcal{D}_j$ is a subset of $\mathcal{N}_j$, confounding variables\footnote{A confounding variable is an unmeasured variable that \red{induces correlation between} the input and output signal of an estimation problem. \cite{Pearl:2000}. \red{A formal definition follows in Definition \ref{def1}}.} can occur in the estimation problem, and these have to be taken into account in the choice of $\mathcal{D}_j$ in order to arrive at consistent estimates of $G_{ji}^0$, see \cite{Dankers&etal_TAC:16}. This situation has been analyzed for uncorrelated disturbances only, i.e. $\Phi_v$ being diagonal.
\item In \cite{Dankers&etal_IFAC:17} relaxed conditions for the previous situation have been formulated, while still staying in the context of MISO identification with $\Phi_v$ being diagonal. This is particularly done by choosing additional predictor input signals that are not in $\mathcal{N}_j$,.i.e. that are no in-neighbors of the output $w_j$ of the target module.
\item Irrespective of noise correlations, an indirect/two-stage identification method can be used to arrive at consistent estimates of $G_{ji}^0$, if particular conditions on $\mathcal{D}_j$ are satisfied, \cite{VandenHof&etal_Autom:13,Dankers&etal_TAC:16}. However the drawback of indirect methods is that they do not allow for a maximum likelihood analysis, i.e. they will not lead to minimum variance results.
\end{itemize}
The step that we would like to make in this paper, is to go beyond consistency properties, and to formulate an identification setup that leads to Maximum Likelihood properties, and thus also minimum variance properties, of the estimated module, for the situation that the disturbance signals can be correlated, i.e. $\Phi_v$ not necessarily being diagonal. This requires a more careful treatment and modelling of the noise that is acting on the different node signals. In \cite{VandenHof&etal_CDC:17} a two-node example network has been studied, which has led to the following two suggestions:
\begin{itemize}
\item confounding variables can be dealt with by modelling correlated disturbances on the node signals, and
\item this can be done by moving from a MISO identification setup to a MIMO setup.
\end{itemize}
These suggestions are being worked out in the current paper, and, as a first step in this analysis, we will stay in the situation of ``full input modeling'', meaning that for every node signal that is included as a predicted output we will include all in-neighbors in the network as predictor input.  A relaxation of this condition is left for future work. We will first present an example to explain the mechanism.

\begin{example}\label{exam1}
Consider the network sketched in Figure \ref{fig1}, and let module $G_{21}$ be the target module for identification.
 If the node signals $w_1$, $w_2$ and $w_4$ can be measured, then a two-input one-output model with inputs $w_1, w_4$ and output $w_2$ will (under the appropriate conditions) lead to a consistent estimate of $G_{21}$ and $G_{24}$, provided that the disturbance signal $v_2$ is uncorrelated to the signals $v_1$ and $v_4$. However if e.g. $v_4$ and $v_2$ are \red{dynamically} correlated, \red{implying that a noise model $H$ of the two-dimensional noise process is non-diagonal}, then consistency is lost for this approach. A solution is then to include $w_4$ in the set of predicted outputs, and by adding node signal $w_3$ as predictor input for $w_4$. We then combine predicting $w_2$ on the basis of $(w_1, w_4)$ with predicting $w_4$ on the basis of $w_3$. The correlation between $v_2$ and $v_4$ is then covered by modelling a $2\times2$ \red{non-diagonal} noise model of the joint process $(v_2, v_4)$.
\end{example}

In the next sections we will formalize the procedure as sketched in Example \ref{exam1} for general networks.

\blue{
\section{Concepts and notation}
\begin{definition}[confounding variable] \label{def1}
	Consider a dynamic network defined by
	\beq
	\label{eqsys1}
	w = Gw + He + r
	\eeq
	with $cov(e) = I$, and consider the graph related to this network, with node signals $w$  and $e$. Let $w_{\smX}$ and $w_{\smY}$ be
	two subsets of measured node signals in $w$, and let $w_{\smZ}$ be the set of unmeasured node signals in $w$.\\
	%Let also $w_{meas}$ and $w_{unmeas}$ be two subsets of node signals in $w$, such that
	%\[
	%w_{\mathcal D} \cup w_{\mathcal Y} \subset w_{meas}.
	%\]
	Then a noise component $e_{\ell}$ in $e$ is a {\it confounding variable for the estimation problem $w_{\smA} \rightarrow w_{\smY}$
		%with measured variables $w_{meas}$}
	}, if in the graph there exist simultaneous paths\footnote{A simultaneous path from $e_1$ to node signal $w_1$ and $w_2$ implies that there exist a path from $e_1$ to $w_1$ as well as from $e_1$ to $w_2$.} from $e_{\ell}$ to node signals $w_{k}, k \in \mA$ and $w_{n}, n \in \mY$, while these paths are either direct\footnote{A direct path from $e_1$ to node signal $w_1$ implies that there exist a path from $e_1$ to $w_1$ which do not pass through nodes in $w$.} or only run through nodes that are not in $w_{\smZ}$. \hfill $\Box$
\end{definition}
}
%\textcolor{red}{
%The important point of this definition, in relation to previous uses, as e.g. in \cite{Dankers&etal_IFAC:17}, is that the driving white noise signals have become part of the set of nodes, so that the correlation structure of the disturbances on the node signals $w$ has become part of the network structure.}

\blue{
We will denote $w_{\smY}$ as the node signals in $w$ that serve as predicted outputs, and $w_{\smD}$ as the node signals in $w$ that serve as predictor inputs. Next we decompose $w_{\smY}$ and $w_{\smD}$ in disjoint sets according to: $ \mY  =  \mQ \cup \{o\} \ ; \ \mD  =  \mQ \cup \mA \cup \mB$ where $w_{\smQ}$ are the node signals that are common in $w_{\smY}$ and $w_{\smD}$; $w_o$ is the output $w_j$ of the target module; if $j \in \mQ$ then $\{o\}$ is void; $\mA \subset \mN_{\smY}$ and $\mB \not\subset \mN_{\smY}$, to be specified later on.
Additionally we denote $w_{\smZ}$ as the node signals in $w$ that are neither predicted output nor predictor input, i.e. $\mZ = \mL \backslash \{ \mD \cup \mY\}$, where $\mL = \{1,2,\cdots L\}$.
%\begin{definition}[confounding variable] \label{def1}
%	Consider a dynamic network defined by
%	\beq
%	\label{eqsys1}
%	w = Gw + He + r
%	\eeq
%	with $cov(e) = I$, and consider the graph related to this network, with node signals $w$  and $e$. Let $\mathcal X \subseteq \mD$ denote the set of indexes of node signals in $w$.\\
%	%Let also $w_{meas}$ and $w_{unmeas}$ be two subsets of node signals in $w$, such that
%	%\[
%	%w_{\mathcal D} \cup w_{\mathcal Y} \subset w_{meas}.
%	%\]
%	Then a noise component $e_{\ell}$ in $e$ is a {\it confounding variable for the estimation problem $w_{\smX} \rightarrow w_{\smY}$
%		%with measured variables $w_{meas}$}
%	}, if in the graph there exist simultaneous paths\footnote{A simultaneous path from $e_1$ to node signal $w_1$ and $w_2$ implies that there exist a path from $e_1$ to $w_1$ as well as from $e_1$ to $w_2$.} from $e_{\ell}$ to node signals in $w_{\smX}$ and $w_{\smY}$, while these paths are either direct\footnote{A direct path from $e_1$ to node signal $w_1$ implies that there exists a path from $e_1$ to $w_1$ which does not pass through nodes in $w$.} or unmeasured \footnote{An unmeasured path is a path that runs through nodes in $w_{\smZ}$ only.}. \hfill $\Box$
%\end{definition}
\begin{figure}[h]
	\centerline{\includegraphics[scale=0.5]{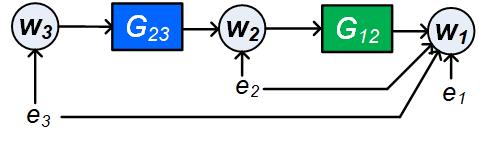}}
	\caption{A simple network with 3 nodes $w_1$, $w_2$, $w_3$ and unmeasured noise sources $e_1$, $e_2$ and $e_3$. $G_{12}$ is the target module to be identified.}
	\label{fig11}
\end{figure}
There can exist two types of confounding variable namely \emph{direct and indirect confounding variable}. For \emph{direct confounding variables} the simultaneous paths mentioned in the definition are both \emph{direct paths}, while in all other cases we refer to the confounding variables as \emph{indirect confounding variables}.
For example, in the network as shown in figure \ref{fig11} with $\mD = \{2\}$, $\mY = \{1\}$ and $\mZ = \{3\}$, for the estimation problem $w_2 \rightarrow w_1$, $e_2$ is a \emph{direct confounding variable} since it has a simultaneous path to $w_1$ and $w_2$ where both the paths are \emph{direct paths}. Meanwhile $e_3$ is an \emph{indirect confounding variable} since it has a simultaneous path to $w_1$ and $w_2$ where one of the path is an unmeasured path\footnote{An unmeasured path is a path that runs through nodes in $w_{\smZ}$ only. Analogously, we can define unmeasured loop through a node $w_i$.}.
}
\section{Algorithm for signal selection: full input case}
In order to arrive at an appropriate identification setup we will take the following strategy:
\begin{itemize}
\item We start by constructing sets $\mQ$ and $\mA$ in such a way that all $w$-in-neighbors of $w_{\smY}$ are included in $w_{\smQ\cup\smA}$ and that all disturbance terms $v_k$, $k\in\mA$ are uncorrelated to disturbance terms $v_{\ell}$, $\ell \in \mY$. In this way we handle the \blue{direct confounding variables}.
\item Then we choose $w_{\smB}$ as a subset of nodes that are not in $w_{\smY}$ nor in $w_{\smA}$. This set needs to be introduced to deal with \blue{the indirect confounding variables}, and will be further specified in Section \ref{sec:main}.
\item Finally, we define the identification setup as the estimation problem $w_{\mathcal D} \rightarrow w_{\mathcal Y}$.
\end{itemize}
The conditions that need to be imposed on the selection of $w_{\smB}$ in order to arrive at attractive properties of the estimation results, will be the main subject of analysis in this paper.

The following algorithm formalizes the procedure as indicated above.

{\bf Algorithm A}
\begin{enumerate}
\item Select target module $G_{ji}$
\item Include $j$ in the index set $\mathcal{Y}$ of node variables that are to be predicted.
%\item Select all $w_k$, $k \in \mathcal{N}_j$ in the set of predictor inputs $\mathcal{U}_j$;
\item For every element $x$ of $\mathcal{Y}$:
\begin{enumerate}
\item For every $k \in \mathcal{N}_x$:
\begin{itemize}
\item include $k$ in $\mD$, and
\item if $v_k$ is correlated with any $w_\ell$, $\ell \in \mathcal{Y}$, then include $k$ in $\mathcal{Y}$;
\end{itemize}
\item If $\mathcal{Y}$ has changed, start step 3 from the beginning again.
\end{enumerate}
\item Determine $\mQ$ as the intersection of $\mY$ and $\mD$;
\item If $j \notin \mQ$ then set $w_o=w_j$, else $w_o$ is void;
\item Determine $\mA = \mD \backslash \mQ$;
\item Make a selection $\mB$ of node signals that are not in $\mY$ and not in $\mA$.
\end{enumerate}

When this algorithm finishes, then the set $\mathcal{Y}$ contains the index set of to be predicted node variables, while for each predicted node variable $x$ in this set, the set of predictor inputs is $\mathcal{N}_x$. In this procedure, input nodes with disturbances that are uncorrelated to output disturbances, will block the further growth of the number of signals in $\mQ$ and $\mA$, while input nodes with correlated disturbances will require further extension of the sets.
%
%
%%%%%%%%%%%%%%%%%%%%%%%%%%%%%%%%%%%%%%%%%%%%%%%%%%%%%%%%%%%%%%%%%%%%%%%%%%%%%%%%%%%%
%\vspace{-0.5cm}
\section {MIMO identification setup}
%
%\vspace{-0.5cm}
On the basis of the decomposition of node signals as defined in the previous section we are going to rewrite the system's equations (\ref{eqsys1}) in the following structured form:
%\vspace{-0.15cm}
\beqr
\label{eq1}
\begin{bmatrix} w_{\smQ} \\ w_o \\ w_{\smB}\\ w_{\smA} \\ w_{\smZ} \end{bmatrix} & = &
\begin{bmatrix} G_{\smQ\smQ} & G_{\smQ o} & G_{\smQ\smB} & G_{\smQ\smA} & G_{\smQ\smZ} \\
                G_{o\smQ} & G_{oo} & G_{o\smB} & G_{o\smA} & G_{o\smZ} \\
                G_{\smB\smQ} & G_{\smB o} & G_{\smB\smB} & G_{\smB\smA} & G_{\smB\smZ} \\
                G_{\smA\smQ} & G_{\smA o} & G_{\smA\smB} & G_{\smA\smA} & G_{\smA\smZ} \\
                G_{\smZ\smQ} & G_{\smZ o} & G_{\smZ\smB} & G_{\smZ\smA} & G_{\smZ\smZ} \end{bmatrix}
                \begin{bmatrix} w_{\smQ} \\ w_o \\ w_{\smB} \\w_{\smA} \\ w_{\smZ} \end{bmatrix} + \nonumber \\
& & + \begin{bmatrix} H_{\smQ\smQ} & H_{\smQ o} & H_{\smQ\smB} & H_{\smQ\smA} & H_{\smQ\smZ} \\
                H_{o\smQ} & H_{oo} & H_{o\smB} & H_{o\smA} & H_{o\smZ} \\
                H_{\smB\smQ} & H_{\smB o} & H_{\smB\smB} & H_{\smB\smA} & H_{\smB\smZ} \\
                H_{\smA\smQ} & H_{\smA o} & H_{\smA\smB} & H_{\smA\smA} & H_{\smA\smZ} \\
                H_{\smZ\smQ} & H_{\smZ o} & H_{\smZ\smB} & H_{\smZ\smA} & H_{\smZ\smZ} \end{bmatrix}
                \begin{bmatrix} e_{\smQ} \\ e_o \\ e_{\smB} \\e_{\smA} \\ e_{\smZ} \end{bmatrix}
\eeqr
%\vspace{-0.3cm}

\noindent where we make the notation agreement that the matrix $H$ is not necessarily monic, and the scaling of the white noise process $e$ is such that $cov(e) = I$. Without loss of generality, we can assume $r = 0$ for the sake of brevity.

If we follow Algorithm A for the signal selection then we satisfy the following assumption.

\begin{assumption}
\label{ass1}
All $w$-in-neighbours of $w_{\smY}$ are collected in $w_{\smQ \cup \smA}$, and all disturbance signals $v_{\smA}$ are uncorrelated to $v_{\smY}$.
\end{assumption}

\begin{proposition}
\label{prop1}
Under the conditions of Assumption \ref{ass1} it follows that in (\ref{eq1}),
%(a) $H_{\smA\smQ}=H_{\smA o}=H_{\smQ\smA}=H_{o\smA}=0 \ \ $
(a) $G_{\smQ\smZ}=G_{o\smZ}=G_{\smQ\smB}=G_{o\smB}=0$;
(b) $G_{oo} = 0$;
(c) If $w_o$ is present then $G_{\smQ o} = 0$.
%\begin{itemize}
%\item[(a)] $H_{\smA\smQ}=H_{\smA o}=H_{\smQ\smA}=H_{o\smA}=0$
%\item[(b)] $G_{\smQ\smZ}=G_{o\smZ}=G_{\smQ\smB}=G_{o\smB}=0$
%\item[(c)] $G_{oo} = 0$
%\item[(d)] If $w_o$ is present then $G_{\smQ o} = 0$ \hfill $\Box$
%\end{itemize}
\end{proposition}
%{\bf Proof}: Part(a) is induced by the condition on decorrelation of disturbance signals. The zeros in the third and fifth column of $G$ are because both $w_{\smB}$ and $w_{\smZ}$ can not contain $w$-in-neighbors of $w_{\smY}$. The zeros in the second column of $G$ are because $w_o$ can not be a predictor input and $G_{oo}$, if present, is scalar.
	{\bf Proof}: The zeros in the third and fifth column of $G$ are because both $w_{\smB}$ and $w_{\smZ}$ can not contain $w$-in-neighbors of $w_{\smY}$. The zeros in the second column of $G$ are because $w_o$ can not be a predictor input and $G_{oo}$, if present, is scalar and hollow. \hfill $\Box$
%\vspace{-0.1cm}
\begin{proposition}
\label{prop2}
Under the conditions of Assumption \ref{ass1}, the system equations for the measured variables $w_{\smD} \cup w_{\smY}$ can be written as
\beqr
\label{eq13}
\begin{bmatrix} w_{\smQ} \\ w_o \\ w_{\smB}\\ w_{\smA}  \end{bmatrix} =
\begin{bmatrix} G_{\smQ\smQ} & 0 & 0 & G_{\smQ\smA}  \\
                G_{o\smQ} & 0 & 0 & G_{o\smA} \\
                \breve G_{\smB\smQ} & \breve G_{\smB o} & \breve G_{\smB\smB} & \breve G_{\smB\smA}  \\
                \breve G_{\smA\smQ} & \breve G_{\smA o} & \breve G_{\smA\smB} & \breve G_{\smA\smA}  \end{bmatrix}
                \begin{bmatrix} w_{\smQ} \\ w_o \\ w_{\smB} \\w_{\smA} \end{bmatrix} +
                \breve v, \nonumber \\
 \breve v = \breve H\begin{bmatrix} e_{\smQ} \\ e_o \\ e_{\smB} \\e_{\smA} \\ e_{\smZ} \end{bmatrix} = \begin{bmatrix} H_{\smQ\smQ} & H_{\smQ o} & H_{\smQ\smB} & H_{\smQ\smA} & H_{\smQ\smZ} \\
                H_{o\smQ} & H_{oo} & H_{o\smB} & H_{o\smA} & H_{o\smZ}  \\
                \breve H_{\smB\smQ} & \breve H_{\smB o} & \breve H_{\smB\smB} & \breve H_{\smB\smA} & \breve H_{\smB\smZ}  \\
                \breve H_{\smA\smQ} & \breve H_{\smA o} & \breve H_{\smA\smB} & \breve H_{\smA\smA} & \breve H_{\smA\smZ} \end{bmatrix}\!\!\!
                \begin{bmatrix} e_{\smQ} \\ e_o \\ e_{\smB} \\e_{\smA} \\ e_{\smZ} \end{bmatrix}
\eeqr
with $cov(e)=I$, and where
\begin{eqnarray}
\breve G_{\smA\star} & = & G_{\smA\star} + G_{\smA z}(I-G_{\smZ\smZ})^{-1}G_{\smZ\star},\\
%\breve H_{\smA\smQ} & = & G_{\smA z}(I-G_{\smZ\smZ})^{-1}H_{\smZ\smQ}, \label{eqhis}\\
%\breve H_{\smA o} & = & G_{\smA\smZ}(I-G_{\smZ\smZ})^{-1}H_{\smZ o},\label{eqhos}\\
%\breve H_{\smA\smB} & = & H_{\smA\smB} + G_{\smA\smZ}(I-G_{\smZ\smZ})^{-1}H_{\smZ\smB},\\
%\breve H_{\smB\smB} & = & H_{\smA\smA} + G_{\smA\smZ}(I-G_{\smZ\smZ})^{-1}H_{\smZ\smA},\\
%\breve H_{\smA\smZ} & = & H_{\smA\smZ} + G_{\smA\smZ}(I-G_{\smZ\smZ})^{-1}H_{\smZ\smZ}, \label{eqhiz}\\
\breve G_{\smB\star} & = & G_{\smB\star} + G_{\smB\smZ}(I-G_{\smZ\smZ})^{-1}G_{\smZ\star},\\
\breve H_{\smA\star} & = & H_{\smA\star} + G_{\smA\smZ}(I-G_{\smZ\smZ})^{-1}H_{\smZ\star}\\
\breve H_{\smB\star} & = & H_{\smB\star} + G_{\smB\smZ}(I-G_{\smZ\smZ})^{-1}H_{\smZ\star}.
\end{eqnarray}
\end{proposition}\hfill $\Box$

{\bf Proof:} See the appendix.

In the sequel we are going to formulate conditions on the choice of node variables in $w_{\smB}$, such that the systems equations for the output variables in $w_{\smY}$ can be written as
\vspace{-0.2cm}
\beq
\label{eq5}
\underbrace{\begin{bmatrix} \!w_{\smQ}\! \\ \!w_o\! \end{bmatrix}}_{w_{\smY}}\!\! =\!\!
\underbrace{\begin{bmatrix} \!\bar G^0_{\smQ\smQ}\! &\! \bar G^0_{\smQ\smB}\! &\! \bar G^0_{\smQ\smA}\!  \\
                \!\bar G^0_{o\smQ}\! & \!\bar G^0_{o\smB}\! &\! \bar G^0_{o\smA}\! \end{bmatrix}}_{\bar G^0}
                \underbrace{\begin{bmatrix} \!w_{\smQ}\! \\ \!w_{\smB}\! \\\!w_{\smA}\! \end{bmatrix}}_{w_{\smD}} +
                \underbrace{\begin{bmatrix} \bar H^0_{\smQ\smQ} & \bar H^0_{\smQ o} \\
                \bar H^0_{o\smQ} & \bar H^0_{oo} \end{bmatrix}}_{\bar H^0}
                \underbrace{\begin{bmatrix} \xi_{\smQ} \\ \xi_o \end{bmatrix}}_{\xi_{\smY}}
\eeq
\vspace{-0.3cm}

\noindent with $\xi_{\smQ}$ and $\xi_o$ white noise processes with dimensions conforming to $w_{\smQ}$ and $w_o$, respectively, with $cov(\xi_{\smY}) = \bar\Lambda$ and with $\bar H^0$ being monic, stable and stably invertible. In the situation of a network system with the system's equations as in (\ref{eq5}) we can set up a predictor model based on a parametrized model set determined by
\[ \M := \left\{(\bar G(\theta), \bar H(\theta), \bar\Lambda(\theta)), \theta\in\Theta\right\}, \]
while the actual data generating system is represented by $\mathcal{S} = (\bar G(\theta_o), \bar H(\theta_o), \bar\Lambda(\theta_0))$.
The corresponding identification problem is defined by considering the one-step-ahead prediction of $w_{\smY}$, according to
\[ \hat w_{\smY}(t|t-1) := \E \{w_{\smY}(t)\ |\ w_{\smY}^{t-1}, w_{\smD}^{t} \} \]
where $w_{\smD}^t$ denotes the past of $w_{\smD}$, i.e. $\{w_{\smD}(k), k\leq t\}$. The resulting prediction error becomes:
\vspace{-0.15cm}
\begin{equation} \label{eqx5}
\varepsilon(t,\theta) := w_{\smY}(t) - \hat w_{\smY}(t|t-1;\theta)
\end{equation}
\[
= \bar H(q,\theta)^{-1}
\left[ w_{\smY}(t) - \bar G(q,\theta)w_{\smD}(t)\right],
\]
and the weighted least squares identification criterion
\vspace{-0.15cm}
\begin{equation}\label{eqx6}
\hat\theta_N = \arg\min_{\theta} \frac{1}{N} \sum_{t=0}^{N-1} \varepsilon^T(t,\theta) W  \varepsilon(t,\theta),
\end{equation}
with $W$ any positive definite weighting matrix. This parameter estimate then leads to an estimated subnetwork $G_{\smY\smD}(q,\hat\theta_N)$, with the estimated target module $G_{ji}(q,\hat\theta_N)$ as a component of this.

\section{Main results}
\label{sec:main}
First we will formulate conditions for the selection of the blocking node variables $w_{\smB}$, that will allow to derive consistent identification results next.
\begin{property}
\label{p1}
Let the node signals $w_{\smB}$ be chosen to satisfy the following properties:
\begin{enumerate}
\item If there are no confounding variables for the estimation problem $w_\mA \rightarrow (w_{\smQ},w_o)$, then $\mB$ is void implying that $w_{\smB}$ is not present;
\item If there are confounding variables for the estimation problem $w_\mA \rightarrow (w_{\smQ},w_o)$, then all of the following conditions are satisfied:
\begin{itemize}
\item[a.] For any confounding variable for the estimation problem $w_\mA \rightarrow (w_{\smQ},w_o)$, the paths from the confounding variable to a node signal $w_{\smA}$ is blocked by a node signal in $w_{\smB}$, where the paths are either direct or unmeasured;
%\footnote{An unmeasured path is a path that runs through nodes in $w_{\smZ}$ only.};
\item[b.] \blue{For every simultaneous path from any $e_{k}$ in $e$ to node signals in $w_{\smB}$ and $w_{\smA}$, at least one of the paths should pass through nodes in $w_{\mL\backslash \mZ}$.}
%For all white noise sources \blue{$e_k$ in $e_{\smQ}$, $e_{o}$, $e_{\smB}$, $e_{\smA}$ and $e_{\smZ}$} there do not exist simultaneous paths from $e_k$ to $w_{\smB}$ and $w_{\smA}$, where the paths are either direct or unmeasured.
Alternatively formulated: the nonmodelled disturbances on $w_{\smB}$ and $w_{\smA}$ are uncorrelated;
\item[c.] There are no direct or unmeasured paths from $w_i$ to node variables in $w_{\smB}$;
\item[d.] There are no direct or unmeasured paths from \blue{$w_j$} to node variables in $w_{\smB}$. \hfill $\Box$
\end{itemize}
\end{enumerate}
\end{property}

Next we can formulate the main consistency result of this paper.

\begin{theorem} \label{theorem1}
Consider a (MIMO) network identification setup with predictor inputs $w_{\smD}$ and predicted outputs $w_{\smY}$, satisfying the conditions of Assumption \ref{ass1} (full input case). Then a prediction error identification method according to (\ref{eqx5})-(\ref{eqx6}), applied to a parametrized model set $\mathcal{M}$ will provide a consistent estimate of the target module $G_{ji}^0$, if
\begin{enumerate}
\item $\mathcal{M}$ is chosen to satisfy $\mathcal{S} \in \mathcal{M}$;
%\item The typical regularity conditions on the data that are required for convergence of the parameter estimate in %prediction error identification, see \cite{Ljung:99}, are satisfied;
\item The blocking node signals $w_{\smB}$ are chosen to satisfy Property \ref{p1};
\item $\Phi_{\kappa}(\omega) > 0 $ for a sufficiently high number of frequencies, where
$ \kappa(t) := \begin{bmatrix} w_{\smD}^\top & \xi_{\smQ}^\top & w_o \end{bmatrix}^\top$;
\item All the elements in $G_{\smQ\smQ}, G_{\smQ\smA}, G_{o\smQ}, G_{o\smA}$ are strictly proper (or) all existing paths/loops from $w_{\smQ}, w_{o}, w_{\smB}$ to $w_{\smQ}$ and from $w_{\smQ}, w_{o}, w_{\smB}$ to $w_{o}$ have at least a delay. \hfill $\Box$
%\hfill $\Box$
\end{enumerate}
\end{theorem}
\medskip
{\bf Proof:} See the appendix.

\medskip
%The detailed proof for the above theorem is provided in \cite{VandenHof&etal:18}.
There are typically two major conditions for arriving at consistency of the target module $G_{ji}$: one needs to be able to deal with the confounding variables through the selection of an appropriate set of (blocking) node variables $w_{\smB}$ that is included as predictor input, and there should be enough excitiaton present in the node signals, which actually reflects a type of identifiability property \cite{Weerts&etal_Autom:18_identifiability}. Note that this excitation condition may require that there are external excitation signals present at some locations, see also \cite{Gevers&Bazanella_CDC:15}. Note that since we are using a direct method for identification, the signals $r$ are not directly used in the predictor model, although they serve the purpose of providing excitation in the network.

\begin{remark}
If we consider the excitation condition formulated in condition 4 of the Theorem \ref{theorem1}, we see a slight variation with respect to the classical condition for closed-loop systems, which typically would contain $w_{\smD}$ and $w_o$ in the vector signal. In the considered network situation where signals can be both input and output, the signal vector in condition 4 is extended with $\xi_{\smQ}$, i.e. the innovation signal related to the disturbances on node signals that are both input and output.
\end{remark}

\medskip
Since in the result of Theorem \ref{theorem1} we arrive at white innovation signals, the result can be extended to formulate Maximum Likelihood properties.

\medskip
\begin{theorem}
\label{theorem2}
Consider the situation of Theorem \ref{theorem1}, and let the conditions for consistency be satisfied. Let $\xi_{\smY}$ be normally distributed, and let $\bar\Lambda(\theta)$ be parametrized independently from $\bar G(\theta)$ and $\bar H(\theta)$. Then, under zero initial conditions, the Maximum Likelihood estimate of $\theta^0$ is
\vspace{-0.15cm} 	
\beqr \label{eq:ML2}
	\hat \theta_N^{ML} & = & \arg \min_{\theta} \det \left ( \frac{1}{N} \sum_{t=1}^N \varepsilon(t,\theta)  \varepsilon^T(t,\theta) \right )
		\\
\Lambda(\hat\theta_N^{ML}) & = & \frac{1}{N} \sum_{t=1}^N  \varepsilon(t,\hat\theta_N^{ML}) \varepsilon^{T}(t,\hat\theta_N^{ML}).
\eeqr
\end{theorem}

{\bf Proof:} Can be shown by following a similar reasoning as in Theorem 1 of \cite{Weerts&etal_Autom:18_reducedrank}. \hfill $\Box$

\section{Examples}

%Walk through the example indicated in Figure \ref{fig1} to show the mechanisms (example is taken from the November 2017 note).

%\begin{figure}[htb]
%\centerline{\includegraphics[scale=0.52]{Network_JD_Exam1.png}}
%	\caption{Example network}
%	\label{fig1}
%\end{figure}

%The target of identification is module $G_{21}$, and all node signals are available for measurements.

%---------------------------------------------------------------------------

In this section we will apply the developed local module identification methodology to two examples of dynamic networks. First we will consider the dynamic network in example \ref{exam1} where $v_2$ and $v_4$ are mutually correlated while the other disturbance signals are uncorrelated with these and with each other. The target of identification is module $G_{21}$, and all node signals are available for measurements. Using the identification method developed in this paper, we first select the signals $w_{\smQ}, w_o, w_{\smA}$ using the algorithm A.
Since $v_2$ and $v_4$ are correlated we choose them both as outputs. Consequently, $w_1$, $w_3$ and $w_4$ are chosen as inputs, so that
\vspace{-0.15cm}
\beqr
  \mY  =  \{2, 4\} & ; & \mD =  \{1, 4, 3\}\\
  \mQ  =  \mY \cap \mD \ =  \{4 \} & ; & \mA  =  \mD \backslash \mQ \ = \{1, 3\} \\
  w_o & = & w_2.
\eeqr
\vspace{-0.65cm}

\noindent For the selection of $w_{\smB}$, according to Property \ref{p1}, we need to check the presence of confounding  variables. Since all disturbance terms $v_k, k \in \mZ \cup \mA$ are uncorrelated to all disturbance terms $v_l, l \in \mY$, there are no confounding variables for the estimation problem $w_\mA \rightarrow (w_{\smQ},w_o)$. Therefore $w_{\smB}$ is void. Now we have the predictor inputs $w_{\smD}$ and the predicted outputs $w_{\smY}$ for the MIMO identification setup that will satisfy the essential conditions of Theorems \ref{theorem1} and \ref{theorem2}.

\begin{example}\label{exam2}
Consider the network sketched in Figure \ref{fig2}, and let module $G_{12}$ be the target module for identification. The disturbance correlation structure in the network is presented in Figure \ref{fig2} with modules in red indicating the noise dynamics.
%\vspace{-0.15cm}
\begin{figure}[htb]
\centerline{\includegraphics[scale=0.45]{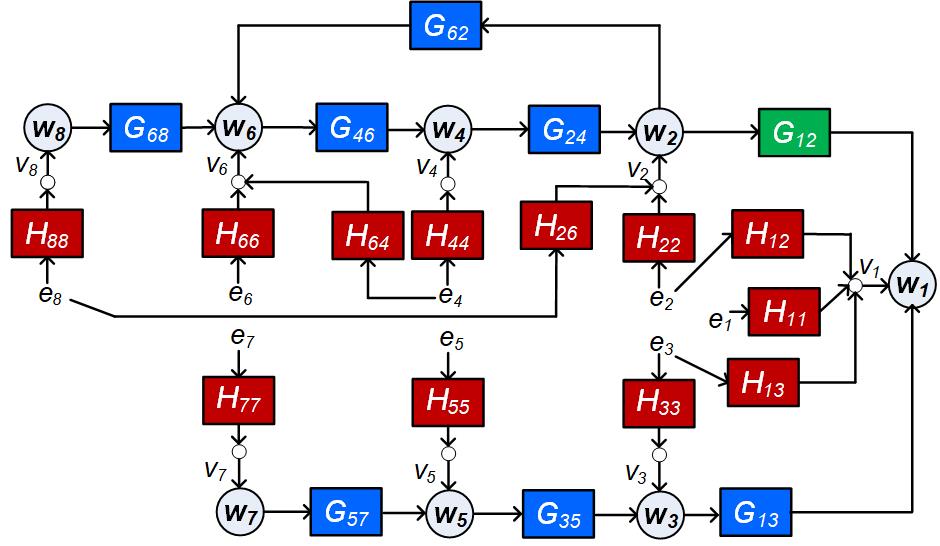}}
\vspace{-0.25cm}
	\caption{Example network}
	\label{fig2}
\end{figure}
%\vspace{-0.25cm}

\noindent The direct method using a MISO predictor, as addressed in \cite{VandenHof&etal_Autom:13}, does not provide a consistent estimate of $G_{12}$ since the disturbance term $v_1$ is correlated with $v_2$ as well as $v_3$ and therefore we resort to the identification framework developed in this paper. Similar to the previous example, the first step will be selection of $w_{\smQ}, w_o, w_{\smA}$ using the algorithm A.

First we select $w_1$ as output and $w_2$ and $w_3$ as inputs. Since $v_2$ and $v_3$ are correlated with $v_1$, both $w_2$ and $w_3$ need to be added as outputs too. Then $w_4$ and $w_5$ need to be added as inputs.
As a result of the first six steps in algorithm A we get,
\vspace{-0.15cm}
\beqr
  \mY  =  \{1, 2, 3\} & ; & \mD =  \{2, 3, 4, 5\}\\
  \mQ  =  \mY \cap \mD \ =  \{2, 3\} & ; & \mA  =  \mD \backslash \mQ \ = \{4, 5\} \\
  w_o & = & w_1.
\eeqr
\vspace{-0.65cm}

\noindent In the resulting situation $e_8$ acts as a confounding variable that affects both input $w_4$ and output $w_2$. As per condition 2a of Property \ref{p1}, the path from $e_8 \rightarrow w_4$ should be blocked by a node signal in $w_{\smB}$, which can be either $w_8$ or $w_6$. In order to choose the node signals $w_{\smB}$, we also need the conditions 2b, 2c and 2d in Property \ref{p1} to be satisfied. $w_6$ cannot be chosen in $w_{\smB}$ since it does not satisfy conditions 2b and 2c in Property 1. The former condition is not satisfied due to the simultaneous path from $e_4$ in $e_{\smA}$ to $w_6$ and $w_4$ and the latter condition is not satisfied due to the path from $w_2$ in $w_i \rightarrow w_6$. When $w_8$ is chosen in $w_{\smB}$, the conditions in Property 1 are satisfied and hence $\mathcal B = \{8\}$. Now we have the predictor inputs $w_{\smD}$ and the predicted outputs $w_{\smY}$ for the MIMO identification setup that provide the consistent and maximum likelihood estimation results of $G_{12}$.
\end{example}

%\section{Discussion}
%
%Possibly use the open-loop cascade example of Giulio (page 3 of the note), to discuss the effect of different choices of parametrized model sets.

\section{CONCLUSIONS}
A new local module identification approach has been presented to identify local modules in a dynamic network with given topology, addressing the situation that process noise on different nodes can be correlated with each other. For this case, it is shown that the problem can be solved by moving from a MISO to a MIMO identification setup.
In this setup the target module is embedded in a MIMO problem with appropriately chosen inputs and outputs, that warrant the consistent estimation of the target module with maximum likelihood properties. \blue{A key part of the procedure is the handling of direct and indirect confounding variables, through the introduction of appropriately chosen additional predictor input node signals (blocking nodes) and predicted output node signals respectively}. We have considered the ``full input'' case, implying that all in-neighbours of an output node are included as input. A further relaxation of this condition is a future step that needs to be made. The presented approach has been illustrated by two examples.

%%%%%%%%%%%%%%%%%%%%%%%%%%%%%%%%%%%%%%%%%%%%%%%%%%%%%%%%%%%%%%%%%%%%%%%%%%%%%%%%%%%%%%%%%%%%%%%
\appendices
\section{Proof for Proposition 2}
From \eqref{eq1}, the fifth (block) row provides the equation for $w_{\smZ}$. Using this equation we can obtain the expression for $G_{\smA\smZ}w_{\smZ}$ and $G_{\smB\smZ}w_{\smZ}$ as,
\begin{equation}
\begin{split}
G&_{\smA\smZ}w_{\smZ} = G_{\smA\smZ}(I-G_{\smZ\smZ})^{-1}[G_{\smZ\smQ}w_{\smQ} + G_{\smZ o}w_o + G_{\smZ\smB}w_{\smB} \\
&+ G_{\smZ\smA}w_{\smA} + H_{\smZ\smQ}e_{\smQ} + H_{\smZ o}e_o + H_{\smZ\smB}e_{\smB} + H_{\smZ\smA}e_{\smA} + H_{\smZ\smZ}e_{\smZ}] \\
G&_{\smB\smZ}w_{\smZ} = G_{\smB\smZ}(I-G_{\smZ\smZ})^{-1}[G_{\smZ\smQ}w_{\smQ} + G_{\smZ o}w_o + G_{\smZ\smB}w_{\smB} \\
&+ G_{\smZ\smA}w_{\smA} + H_{\smZ\smQ}e_{\smQ} + H_{\smZ o}e_o + H_{\smZ\smB}e_{\smB} + H_{\smZ\smA}e_{\smA} + H_{\smZ\smZ}e_{\smZ}]
\end{split}
\end{equation}
Substituting the above expression for $G_{\smB\smZ}w_{\smZ}$ and $G_{\smA\smZ}w_{\smZ}$ in the third and fourth (block) row of \eqref{eq1} respectively, we obtain the result of the proposition \hfill $\Box$

\section{Proof for Theorem 1}
%Before presenting the proof of Theorem 1, we present certain lemmas and propositions which specifies the need for condition 3 in Theorem \ref{theorem1} to be satisfied.
%
In order to prove the Theorem 1, a few preparatory results need to be derived which are given below.

\begin{lemma}
	\label{lemma1}
	Consider a dynamic network as defined in (\ref{eq13}) and consider a white noise source $e_x$ and two node signals $w_1$ and $w_2$.
%	There does not exist simultaneous paths from $e_x$ to $w_1$ and to $w_2$ that are either direct or through unmeasured nodes, only if
	For every simultaneous path from $e_x$ to $w_1$ and to $w_2$, at least one of the paths pass through nodes $w_{\mL\backslash \mZ}$, only if
	\[ H_{1x} H_{2x}^* = 0, \]
	where $H_{1x}$, $H_{2x}$ are the noise model transfers from $e_x$ to $w_1$, $w_2$ respectively.
\end{lemma}
{\bf Proof:} $w_1$ and $w_2$ are correlated through $e_x$ only if $\Phi_{w_1e_x} \cdot \Phi_{e_xw_2}$ is unequal to zero. This matrix has a dimension of $dim(w_1)\ \times dim(w_2)$. From the spectrum expression the result of the Lemma follows directly. \hfill $\Box$\\

%%%%%%%%%%%%%%%%%%%%%%%%%%%%%%%%%%%%%%%%%%%%%%%%%%%%%%%%%%%%%%%%%%%%%%%%%%%%%%%%%%%%%%
\begin{proposition}
	\label{prop12}
	Consider the situation of Proposition \ref{prop2}. If conditions 2a and 2b in Property \ref{p1} are satisfied then the spectral density $\Phi_{\breve{v}}$ has the unique spectral factorization
	$\Phi_{\breve v} = \tilde H\Lambda\tilde H^*$ where $\Lambda$ can be any matrix and $\tilde H$ is monic, stable, minimum phase, and of the form
	\beq
	\label{eq20}
	\tilde H = \begin{bmatrix}
		\tilde{H}_{11} & \tilde{H}_{12} & \tilde{H}_{13} & 0\\
		\tilde{H}_{21} & \tilde{H}_{22} & \tilde{H}_{23} & 0\\
		\tilde{H}_{31} & \tilde{H}_{32} & \tilde{H}_{33} & 0\\
		0 & 0 & 0 & \tilde{H}_{44}\\
	\end{bmatrix}
	\eeq
	where the block dimensions are conformable to the dimensions of $w_{\smQ}$, $w_o$, $w_{\smB}$ and $w_{\smA}$ respectively.
\end{proposition}

{\bf Proof:}
Starting from the expression (\ref{eq13}) the spectral density $\Phi_{\breve v}$ can be written as $\breve H\breve H^*$
while it is denoted as
\beq\label{eq32x}
\Phi_{\breve v} = \begin{bmatrix} \Phi_{\breve v_{\smQ}} & \Phi_{\breve v_{\smQ}\breve v_o} & \Phi_{\breve v_{\smQ}\breve v_{\smB}} & \Phi_{\breve v_{\smQ}\breve v_{\smA}}\\
	\Phi_{\breve v_o\breve v_{\smQ}} & \Phi_{\breve v_o} & \Phi_{\breve v_o\breve v_{\smB}} & \Phi_{\breve v_o\breve v_{\smA}} \\
	\Phi_{\breve v_{\smB}\breve v_{\smQ}} & \Phi_{\breve v_{\smB}\breve v_o} & \Phi_{\breve v_{\smB}} & \Phi_{\breve v_{\smB}\breve v_{\smA}} \\
	\Phi_{\breve v_{\smA}\breve v_{\smQ}} & \Phi_{\breve v_{\smA}\breve v_o} & \Phi_{\breve v_{\smA}\breve v_{\smB}} & \Phi_{\breve v_{\smA}} \end{bmatrix}.
\eeq
In this structure we are particularly going to analyse the elements
\beq
\resizebox{1.0\hsize}{!}{$
	\begin{split}
	\Phi_{\breve v_{\smQ}\breve v_{\smA}} = & \breve H_{\smQ\smQ}\breve H_{\smA\smQ}^* + \breve H_{\smQ o}\breve H_{\smA o}^* + \breve H_{\smQ\smB}\breve H_{\smA\smB}^*+ \breve H_{\smQ\smA}\breve H_{\smA\smA}^*+ \breve H_{\smQ\smZ}\breve H_{\smA\smZ}^* \\
	\Phi_{\breve v_o\breve v_{\smA}} = & \breve H_{o\smQ}\breve H_{\smA\smQ}^* + \breve H_{oo}\breve H_{\smA o}^* + \breve H_{o\smB}\breve H_{\smA\smB}^*+ \breve H_{o\smA}\breve H_{\smA\smA}^* +\breve H_{o\smZ}\breve H_{\smA\smZ}^*\\
	\Phi_{\breve v_{\smB}\breve v_{\smA}} = & \breve H_{\smB\smQ}\breve H_{\smA\smQ}^* + \breve H_{\smB o}\breve H_{\smA o}^* + \breve H_{\smB\smB}\breve H_{\smA\smB}^* + \breve H_{\smB\smA}\breve H_{\smA\smA}^* + \breve H_{\smB\smZ}\breve H_{\smA\smZ}^* \label{eqfbi}
	\end{split}$
}
\eeq
In order to arrive at the block diagonal structure for $\tilde H$ as mentioned in the Proposition, we need to arrive at a similar block diagonal structure of $\Phi_{\breve v}$, and therefore the three terms listed above, need to be shown to be equal to $0$. \\
We have $\breve H_{\smA\star} = H_{\smA\star} + G_{\smA\smZ}(I-G_{\smZ\smZ})^{-1}H_{\smZ\star} = H_{\smA\star} + H_{\smA\star}^{(i)}$. Here $H_{\smA\star}^{(i)}$ includes the transfer in the path from $e_{\star}$ to $w_{\smA}$ through nodes in $w_{\smZ}$. Similarly we can write $\breve H_{\smQ\star}, \breve H_{o\star}, \breve H_{\smB\star}$. Rewriting the first two equations in \eqref{eqfbi} as,
\beq
\begin{split}\label{eq43xx}
	\Phi_{\breve v_{\smQ}\breve v_{\smA}} &= \sum_{\star = \mQ, o, \mA, \mB, \mZ}H_{\smQ\star}H_{\smA\star}^* + \sum_{\star = \mQ, o, \mA, \mB, \mZ}H_{\smQ\star}H_{\smA\star}^{(i)*} \\ &+ \sum_{\star = \mQ, o, \mA, \mB, \mZ}H_{\smQ\star}^{(i)*}H_{\smA\star}^* + \sum_{\star = \mQ, o, \mA, \mB, \mZ}H_{\smQ\star}^{(i)*}H_{\smA\star}^{(i)*}
\end{split}
\eeq
Similarly we can write $\Phi_{\breve v_o\breve v_{\smA}}$. If Assumption \ref{ass1} is satisfied, $v_{\smQ}$ is uncorrelated with $v_{\smA}$ and $v_{o}$ is uncorrelated with $v_{\smA}$. Therefore $\Phi_{v_{\smQ}v_{\smA}} = \sum_{\star = \mQ, o, \mA, \mB, \mZ}H_{\smQ\star}H_{\smA\star}^* = 0$ and $\Phi_{v_{o}v_{\smA}} = \sum_{\star = \mQ, o, \mA, \mB, \mZ}H_{o\star}H_{\smA\star}^* = 0$. If condition 2a in Property \ref{p1} is satisfied, the condition of Lemma \ref{lemma1} then implies that $H_{\smQ\star}H_{\smA\star}^{(i)*} = H_{o\star}H_{\smA\star}^{(i)*} = H_{\smQ\star}^{(i)*}H_{\smA\star} = H_{o\star}^{(i)*}H_{\smA\star} = H_{\smQ\star}^{(i)*}H_{\smA\star}^{(i)*} = H_{o\star}^{(i)*}H_{\smA\star}^{(i)*} = 0$ for $\star = \mQ, o, \mA, \mB, \mZ$. Therefore the latter part of the sum in \eqref{eq43xx} becomes 0 and the total sum is also zero which implies that $\Phi_{\breve v_{\smQ}\breve v_{\smA}} = \Phi_{\breve v_o\breve v_{\smA}} = 0$. If condition 2b in Property \ref{p1} is satisfied, directly implying from lemma \ref{lemma1} we have $\Phi_{\breve v_{\smB}\breve v_{\smA}} = 0$.

As a result we can write the spectrum in equation \eqref{eq32x} as,
\beq\label{eq34}
\Phi_{\breve v} = \begin{bmatrix} \Phi_{\breve v_{\smQ}} & \Phi_{\breve v_{\smQ}\breve v_o} & \Phi_{\breve v_{\smQ}\breve v_{\smB}} & 0\\
	\Phi_{\breve v_o\breve v_{\smQ}} & \Phi_{\breve v_o} & \Phi_{\breve v_o\breve v_{\smB}} & 0 \\
	\Phi_{\breve v_{\smB}\breve v_{\smQ}} & \Phi_{\breve v_{\smB}\breve v_o} & \Phi_{\breve v_{\smB}} & 0 \\
	0 & 0 & 0 & \Phi_{\breve v_{\smA}} \end{bmatrix}
\eeq
Then the spectral density $\Phi_{\breve{v}}$ has the unique spectral factorization
\[
\Phi_{\breve v} = \begin{bmatrix}
F_{11}\Lambda_{11}F_{11}^* & 0 \\ 0 & F_{22}\Lambda_{22}F_{22}^*
\end{bmatrix}= \tilde H\Lambda\tilde H^*
\]
where $\tilde H$ is monic, stable, minimum phase and of the form given in proposition \ref{prop12} \hfill $\Box$\\
%%%%%%%%%%%%%%%%%%%%%%%%%%%%%%%%%%%%%%%%%%%%%%%%%%%%%%%%%%%%%%%%%%%%%%%%%%%%%%%%%%%%%%

\begin{proposition}
	\label{prop5a}
	In the situation of Proposition \ref{prop12}, the network representation (\ref{eq13}) can be equivalently transformed to the representation
	\beq
	\label{eq194}
	\begin{split}
	\begin{bmatrix} \! w_{\smQ} \! \\ \! w_o \! \\ \! w_{\smB} \! \\ \! w_{\smA} \! \end{bmatrix} \!\!\! = \!\! &
	\begin{bmatrix} \! \breve G_{\smQ\smQ}' \! & \! 0 \! & \! \breve G_{\smQ\smB}' \! & \! \breve G_{\smQ\smA}'\!\\
		\! \breve G_{o\smQ} \! & \! 0 \! & \! \breve G_{o\smB} \! & \! \breve G_{o\smA} \!\\
		\! \breve G_{\smB\smQ} \! & \! \breve G_{\smB o} \! & \! \breve G_{\smB\smB} \! & \! \breve G_{\smB\smA} \! \\
		\! \breve G_{\smA\smQ} \! & \! \breve G_{\smA o} \! & \! \breve G_{\smA\smB} \! & \! \breve G_{\smA\smA} \!\end{bmatrix} \!\!\!
	\begin{bmatrix} \! w_{\smQ} \! \\ \! w_o \! \\ \! w_{\smB} \! \\ \! w_{\smA} \!   \end{bmatrix} \!\!\!
	+ \!\!\! \begin{bmatrix}
		\! \tilde{H}_{11}'' \!\! & \!\! \tilde{H}_{12}'' \!\! & \!\! 0 \!\! & \!\! 0 \!\\
		\! \tilde{H}_{21}' \!\! & \!\! \tilde{H}_{22}' \!\! & \!\! 0 \!\! & \!\! 0 \!\\
		\! \tilde{H}_{31} \!\! & \!\! \tilde{H}_{32} \!\! & \!\! \tilde{H}_{33} \!\! & \!\! 0 \! \\
		\! 0 \!\! & \!\! 0 \!\! & \!\! 0 \!\! & \!\! \tilde{H}_{44} \! \\
	\end{bmatrix}\!\!\!
	\begin{bmatrix}
			\xi_{\smQ} \\
			\xi_o \\
			\xi_{\smB} \\
			\xi_{\smA}
	\end{bmatrix}
\end{split}
	\eeq
	where
	\beqr
	\label{eq1941}
	\breve G_{o\smQ} & = & (1+\tilde{H}_{23}\tilde{H}_{33}^{-1}\breve G_{\smB o})^{-1}(G_{o\smQ}-\tilde{H}_{23}\tilde{H}_{33}^{-1}\breve G_{\smB\smQ}),\\
	\breve G_{o\smB} & = & (1+\tilde{H}_{23}\tilde{H}_{33}^{-1}\breve G_{\smB o})^{-1}(\tilde{H}_{23}\tilde{H}_{33}^{-1}(I - \breve G_{\smB\smB})),\\
	\label{eq1942}
	\breve G_{o\smA} & = & (1+\tilde{H}_{23}\tilde{H}_{33}^{-1}\breve G_{\smB o})^{-1}(G_{o\smA}-\tilde{H}_{23}\tilde{H}_{33}^{-1}\breve G_{\smB\smA}),\\
	\tilde{H}_{21}' & = & (1+\tilde{H}_{23}\tilde{H}_{33}^{-1}\breve G_{\smB o})^{-1}(\tilde{H}_{21}-\tilde{H}_{23}\tilde{H}_{33}^{-1}\tilde H_{31}),\\
	\tilde{H}_{22}' & = & (1+\tilde{H}_{23}\tilde{H}_{33}^{-1}\breve G_{\smB o})^{-1}(\tilde{H}_{22}-\tilde{H}_{23}\tilde{H}_{33}^{-1}\tilde H_{32})\\
	\breve G_{\smQ\smQ} & = &  G_{\smQ\smQ}-\tilde{H}_{13}\tilde{H}_{33}^{-1}(\breve G_{\smB\smQ} + \breve G_{\smB o}\breve G_{o\smQ}),\\
	\breve G_{\smQ\smA} & = & G_{\smQ\smA}-\tilde{H}_{13}\tilde{H}_{33}^{-1}(\breve G_{\smB\smA} + \breve G_{\smB o}\breve G_{o\smA}),\\
	\breve G_{\smQ\smB} & = & \tilde{H}_{13}\tilde{H}_{33}^{-1}(I - \breve G_{\smB\smB} - \breve G_{\smB o}\breve G_{o\smB}),\\
	\tilde{H}_{11}' & = & \tilde{H}_{11}-\tilde{H}_{13}\tilde{H}_{33}^{-1}(\tilde H_{31} + \breve G_{\smB o}\tilde H_{21}'),\\
	\tilde{H}_{12}' & = & \tilde{H}_{12}-\tilde{H}_{13}\tilde{H}_{33}^{-1}(\tilde H_{32} + \breve G_{\smB o}\tilde H_{22}'),\\
\label{eq1943}
\breve G_{\smQ\smQ}' & = & (I - \mathrm{diag}(\breve G_{\smQ\smQ}))^{-1}(\breve G_{\smQ\smQ} - \mathrm{diag}(\breve G_{\smQ\smQ})) ,\\
\label{eq1944}
\breve G_{\smQ\smA}' & = & (I - \mathrm{diag}(\breve G_{\smQ\smQ}))^{-1}\breve G_{\smQ\smA},\\\label{eq22}
\breve G_{\smQ\smB}' & = & (I - \mathrm{diag}(\breve G_{\smQ\smQ}))^{-1}\breve G_{\smQ\smB},\\
\tilde H_{11}'' & = & (I - \mathrm{diag}(\breve G_{\smQ\smQ}))^{-1}\tilde H_{11}',\\
\tilde H_{12}'' & = & (I - \mathrm{diag}(\breve G_{\smQ\smQ}))^{-1}\tilde H_{12}'.
%	\ \ \ \ \ \ \ \ \ \ \ \ \Box
	\eeqr
	%and
	%\[
	%\begin{bmatrix}
	%\tilde{H}_{11}' & \tilde{H}_{12}' \\
	%\tilde{H}_{21}' & \tilde{H}_{22}'
	%\end{bmatrix}
	%\]
	%is monic.
\end{proposition}
{\bf Proof:} Resulting from proposition \ref{prop12} we can write the network representation \eqref{eq13} as,
%\beq
%\label{eq18}
%\begin{bmatrix} w_s \\ w_o \\ w_b\\ w_i \end{bmatrix} =
%\begin{bmatrix} G_{ss} & 0 & 0 & G_{si}\\
%	G_{os} & 0 & 0 & G_{oi}\\
%	\breve G_{bs} & \breve G_{bo} & \breve G_{bb} & \breve G_{bi} \\
%	\breve G_{is} & \breve G_{io} & \breve G_{ib} & \breve G_{ii} \end{bmatrix}
%\begin{bmatrix} w_s \\ w_o \\ w_b \\ w_i  \end{bmatrix}
%+ \begin{bmatrix}
%	\tilde{H}_{11} & \tilde{H}_{12} & \tilde{H}_{13} & 0\\
%	\tilde{H}_{21} & \tilde{H}_{22} & \tilde{H}_{23} & 0\\
%	\tilde{H}_{31} & \tilde{H}_{32} & \tilde{H}_{33} & 0\\
%	0 & 0 & 0 & \tilde{H}_{44}\\
%\end{bmatrix}
%\begin{bmatrix} \xi_s \\ \xi_o \\ \xi_b \\ \xi_i \end{bmatrix}
%\eeq
\beq
\label{eq18}
\begin{bmatrix} w_{\smQ} \\ w_o \\ w_{\smB}\\ w_{\smA} \end{bmatrix} =
\begin{bmatrix} G_{\smQ\smQ} & 0 & 0 & G_{\smQ\smA}\\
	G_{o\smQ} & 0 & 0 & G_{o\smA}\\
	\breve G_{\smB\smQ} & \breve G_{\smB o} & \breve G_{\smB\smB} & \breve G_{\smB\smA} \\
	\breve G_{\smA\smQ} & \breve G_{\smA o} & \breve G_{\smA\smB} & \breve G_{\smA\smA} \end{bmatrix}
\begin{bmatrix} w_{\smQ} \\ w_o \\ w_{\smB} \\ w_{\smA}  \end{bmatrix}
+ \tilde H
\begin{bmatrix} \xi_{\smQ} \\ \xi_o \\ \xi_{\smB} \\ \xi_{\smA} \end{bmatrix}
\eeq
Pre-multiplying equation \eqref{eq18} by
\beq
\begin{bmatrix}
	I & 0 & -\tilde{H}_{13}\tilde{H}_{33}^{-1} & 0\\
	0 & 1 & -\tilde{H}_{23}\tilde{H}_{33}^{-1} & 0\\
	0 & 0 & I & 0\\
	0 & 0 & 0 & I\\
\end{bmatrix}
\eeq
and moving the terms dependent on $w_{\smB}$ on left side to the right side of the equation, we get
\beqs
\label{eq19.3}
	\begin{split}
	\begin{bmatrix} \!\! w_{\smQ}\!\! \\ \!\!w_o\!\! \\ \!\!w_{\smB}\!\! \\ \!\!w_{\smA}\!\! \end{bmatrix} \!\!\! & = \!\!
	\begin{bmatrix} \! \check G_{\smQ\smQ} \!\! & \!\!\check G_{\smQ o}\!\! & \!\!\check G_{\smQ\smB}\!\! & \!\!\check G_{\smQ\smA}\!\\
	\!\check G_{o\smQ}\!\! & \!\!\check G_{oo}\!\! & \!\!\check G_{o\smB}\!\! & \!\!\check G_{o\smA}\!\\
	\!\breve G_{\smB\smQ} \!\!& \!\!\breve G_{\smB o} \!\!& \!\!\breve G_{\smB\smB}\!\! &\!\! \breve G_{\smB\smA}\! \\
	\!\breve G_{\smA\smQ}\!\! & \!\!\breve G_{\smA o}\!\! & \!\!\breve G_{\smA\smB}\!\! & \!\!\breve G_{\smA\smA}\! \end{bmatrix}\!\!\!
	\begin{bmatrix} \!\!w_{\smQ}\!\! \\ \!\!w_o\!\! \\ \!\!w_{\smB}\!\! \\ \!\!w_{\smA}\!\!  \end{bmatrix} \!\!\!+\!\!\! \begin{bmatrix}
	\!\check{H}_{11}\!\! &\!\! \check{H}_{12}\!\! & \!\!0\!\! &\!\! 0\!\\
	\!\check{H}_{21}\!\! &\!\! \check{H}_{22}\!\! &\!\! 0\!\! &\!\! 0\!\\
	\!\tilde{H}_{31}\!\! & \!\!\tilde{H}_{32}\!\! &\!\! \tilde{H}_{33}\!\! &\!\! 0\!\\
	\!0\!\! &\!\! 0 \!\!& \!\!0 \!\!&\!\! \tilde{H}_{44}\!\\
	\end{bmatrix}\!\!\!
	\begin{bmatrix}
	\!\!\xi_{\smQ}\!\!\\
	\!\!\xi_o\!\! \\
	\!\!\xi_{\smB}\!\! \\
	\!\!\xi_{\smA}\!\!
	\end{bmatrix}
	\end{split}
\eeqs
where $\check G_{\smQ \smQ}\!\!\!=\!\!\! G_{\smQ\smQ}\!\!\!-\!\!\!\tilde{H}_{13}\tilde{H}_{33}^{-1}\breve G_{\smB\smQ}$, $\check G_{\smQ o}\!\!\! =\!\!\! -\tilde{H}_{13}\tilde{H}_{33}^{-1}\breve G_{\smB o}$, $\check G_{\smQ\smB}\!\!\! =\!\!\! \tilde{H}_{13}\tilde{H}_{33}^{-1}(I - \breve G_{\smB\smB})$, $\check G_{\smQ\smA} \!\!\!=\!\!\! G_{\smQ\smA}\!\!\!-\!\!\!\tilde{H}_{13}\tilde{H}_{33}^{-1}\breve G_{\smB\smA}$, $\check G_{o\smQ} \!\!\!=\!\!\! G_{o\smQ}\!\!\!-\!\!\!\tilde{H}_{23}\tilde{H}_{33}^{-1}\breve G_{\smB\smQ}$, $\check G_{oo} \!\!\!=\!\!\! -\tilde{H}_{23}\tilde{H}_{33}^{-1}\breve G_{\smB o}$, $\check G_{o\smB}\!\!=\!\!\tilde{H}_{23}\tilde{H}_{33}^{-1}(I - \breve G_{\smB\smB})$, $\check G_{o\smA}\!\!=\!\!G_{o\smA}\!\!-\!\!\tilde{H}_{23}\tilde{H}_{33}^{-1}\breve G_{\smB\smA}$, $\check{H}_{11}\!\!=\!\!\tilde{H}_{11}\!\!-\!\!\tilde{H}_{13}\tilde{H}_{33}^{-1}\tilde H_{31}$, $\check{H}_{12}\!\!=\!\!\tilde{H}_{12}\!\!-\!\!\tilde{H}_{13}\tilde{H}_{33}^{-1}\tilde H_{32}$, $\check{H}_{21}\!\!=\!\!\tilde{H}_{21}\!\!-\!\!\tilde{H}_{23}\tilde{H}_{33}^{-1}\tilde H_{31}$, $\check{H}_{22}\!\!=\!\!\tilde{H}_{22}\!\!-\!\!\tilde{H}_{23}\tilde{H}_{33}^{-1}\tilde H_{32}$.
%\beq
%\label{eq19.3}
%\resizebox{1.0\hsize}{!}{$
%\begin{split}
%	\begin{bmatrix} \! w_{\smQ}\! \\ \!w_o\! \\ \!w_{\smB}\! \\ \!w_{\smA}\! \end{bmatrix} \!\!\! & = \!\!
%	\begin{bmatrix} \! G_{\smQ\smQ}-\tilde{H}_{13}\tilde{H}_{33}^{-1}\breve G_{\smB\smQ} \!\! & \!\!-\tilde{H}_{13}\tilde{H}_{33}^{-1}\breve G_{\smB o}\!\! & \!\!\tilde{H}_{13}\tilde{H}_{33}^{-1}(I - \breve G_{\smB\smB})\!\! & \!\!G_{\smQ\smA}-\tilde{H}_{13}\tilde{H}_{33}^{-1}\breve G_{\smB\smA}\!\\
%		\!G_{o\smQ}-\tilde{H}_{23}\tilde{H}_{33}^{-1}\breve G_{\smB\smQ}\!\! & \!\!-\tilde{H}_{23}\tilde{H}_{33}^{-1}\breve G_{\smB o}\!\! & \!\!\tilde{H}_{23}\tilde{H}_{33}^{-1}(I - \breve G_{\smB\smB})\!\! & \!\!G_{o\smA}-\tilde{H}_{23}\tilde{H}_{33}^{-1}\breve G_{\smB\smA}\!\\
%		\!\breve G_{\smB\smQ} \!\!& \!\!\breve G_{\smB o} \!\!& \!\!\breve G_{\smB\smB}\!\! &\!\! \breve G_{\smB\smA}\! \\
%		\!\breve G_{\smA\smQ}\!\! & \!\!\breve G_{\smA o}\!\! & \!\!\breve G_{\smA\smB}\!\! & \!\!\breve G_{\smA\smA}\! \end{bmatrix}\!\!\!
%	\begin{bmatrix} \!w_{\smQ}\! \\ \!w_o\! \\ \!w_{\smB}\! \\ \!w_{\smA}\!  \end{bmatrix}\\
%	&+ \begin{bmatrix}
%		\tilde{H}_{11}-\tilde{H}_{13}\tilde{H}_{33}^{-1}\tilde H_{31} & \tilde{H}_{12}-\tilde{H}_{13}\tilde{H}_{33}^{-1}\tilde H_{32} & 0 & 0\\
%		\tilde{H}_{21}-\tilde{H}_{23}\tilde{H}_{33}^{-1}\tilde H_{31} & \tilde{H}_{22}-\tilde{H}_{23}\tilde{H}_{33}^{-1}\tilde H_{32} & 0 & 0\\
%		\tilde{H}_{31} & \tilde{H}_{32} & \tilde{H}_{33} & 0\\
%		0 & 0 & 0 & \tilde{H}_{44}\\
%	\end{bmatrix}
%	\begin{bmatrix}
%		\xi_{\smQ} \\
%		\xi_o \\
%		\xi_{\smB} \\
%		\xi_{\smA}
%	\end{bmatrix}
%\end{split}$
%}
%\eeq
Since the (1,2) and (2,2) block-elements of $G$ have become unequal to $0$ now, the structure does not comply anymore to the required identification structure in (\ref{eq5}), we have to clear these two elements by variable substitution. We follow the following steps sequentially:
\begin{itemize}
\item For the second row of the equation we bring the $w_o$-dependent terms to the left side, and multiply the row with the inverse of the matrix appearing there;
\item The (1,2) block element in the $G$-matrix can be removed by subsituting $w_0$ from the second row into the first row, leading to expression in equation \eqref{eq194}.
\end{itemize}
Now we have the (1,1) block-elements of the resulting matrix to have non-zero elements in the diagonal. By multiplying the first row with a diagonal matrix $(I - \mathrm{diag}(\breve G_{\smQ\smQ}'))^{-1}$, we lead to expression in equation \eqref{eq194} where $\breve G_{\smQ\smQ}''$ has diagonal elements as zero. \hfill $\Box$

\medskip
We have now arrived at the system description (\ref{eq194}) and by extracting from (\ref{eq194}) the expression
	\beq
	\label{eqxx5}
	\begin{bmatrix} \! w_{\smQ} \!\\ \!w_o\! \end{bmatrix}  = \underbrace{\begin{bmatrix} \breve G_{\smQ\smQ}' & \breve G_{\smQ\smB}' & \breve G_{\smQ\smA}'\\
		\breve G_{o\smQ} & \breve G_{o\smB} & \breve G_{o\smA} \end{bmatrix}}_{\breve G}
	\!\!\!\begin{bmatrix} w_{\smQ} \\ w_{\smB} \\ w_{\smA}  \end{bmatrix}
	+\begin{bmatrix}
		\tilde{H}_{11}' & \tilde{H}_{12}' \\
		\tilde{H}_{21}' & \tilde{H}_{22}'
	\end{bmatrix}\!\!\!
	\begin{bmatrix}
			\xi_{\smQ} \\
			\xi_o
	\end{bmatrix}
	\eeq
	we obtain an expression that satisfies the structure of \eqref{eq5} with $\bar G_{\smQ\smQ}^0$ hollow i.e. the diagonal elements are zero. If, on the basis of this equation, the elements of $\breve G$ in \eqref{eqxx5} can be estimated consistently, then the question that remains left is, under which conditions do these elements $\breve G$ reflect the actual target module that we would like to identify. This is addressed next.\\
The target module that is the objective of our identification is given by $G_{ji}^0$, with $w_j \in (w_{\smQ},w_o)$ and $w_i \in (w_{\smA},w_{\smQ})$.	Now there are two situations that we can distinguish: either $w_o = w_j$ (or) $w_o$ is void and $w_j \in w_s$.
	For both situations we formulate the conditions that guarantee that the target module remains invariant in $\breve G$.
	\begin{proposition}
		\label{prop5b}
		Consider the situation of Proposition \ref{prop5a}, and consider the target module $G_{ji}=G_{oi}$. Then $
		\breve G_{oi} = G_{oi}
		$
		if condition 2c and condition 2d of Property \ref{p1} are satisfied.
	\end{proposition}
{\bf Proof:} The target module is a module in $\breve G_{o\smQ}$ or $\breve G_{o\smA}$. The expression for $\breve G_{o\smQ}$ and $\breve G_{o\smA}$ is given in \eqref{eq1941} and \eqref{eq1942} respectively.
%
%$\breve G_{os}$, $\breve G_{oi}$ are given by (\ref{eqgos}) and (\ref{eqgoi}) as,
%\begin{equation} \label{eq60}
%\begin{split}
%\breve G_{os} &= (1+\tilde{H}_{23}\tilde{H}_{33}^{-1}\breve G_{bo})^{-1}(G_{os}-\tilde{H}_{23}\tilde{H}_{33}^{-1}\breve G_{bs})\\
%%\breve G_{ob} &= (1+\tilde{H}_{23}\tilde{H}_{33}^{-1}\breve %G_{bo})^{-1}(\tilde{H}_{23}\tilde{H}_{33}^{-1}(I - \breve G_{bb}))\\
%\breve G_{oi} &= (1+\tilde{H}_{23}\tilde{H}_{33}^{-1}\breve G_{bo})^{-1}(G_{oi}-\tilde{H}_{23}\tilde{H}_{33}^{-1}\breve G_{bi})
%\end{split}
%\end{equation}
%
Now it can be observed that for the target module to be invariant, it is sufficient to require that $\breve G_{\smB o} = 0$ and $\breve G_{\smB i} = 0$. This holds irrespective whether $w_i$ is in $w_{\smQ}$ or in $w_{\smA}$. $\breve G_{\smB i} = 0$ and $\breve G_{\smB o} = 0$ are achieved when condition 2c and condition 2d in Property \ref{p1} are satisfied respectively. \hfill $\Box$\\
%%%%%%%%
%\textcolor{blue}{If the output of the target module $w_j$ is also used as a predictor input, then $w_j$ is included in $w_s$ and $w_o$ is void. This situation is considered next.}
\begin{proposition}
	\label{prop5c}
	Consider the situation of Proposition \ref{prop5a} and let $w_o$ be void. Then the target module remains invariant
%$
%	\breve G_{\smQ i}' = G_{\smQ i}
%	%  ,\ \ \ \mbox{and} \ \
%	%  \breve G_{oi_t} & = G_{oi_t}
%$
	if condition 2c and condition 2d of Property \ref{p1} are satisfied.
\end{proposition}
{\bf Proof:} The target module is a module in $\breve G_{\smQ\smQ}'$ or $\breve G_{\smQ\smA}'$. The expression for $\breve G_{\smQ\smQ}'$ and $\breve G_{\smQ\smA}'$ is given in \eqref{eq1943} and \eqref{eq1944} respectively. But since $w_o$ is void, the expression will be void of terms dependent on $w_o$.  Consider the target module $G_{ji}=G_{\smQ_o i}$. Following the same reasoning as in proposition \ref{prop5b} (when condition 2c and condition 2d of Property \ref{p1} are satisfied), it can be proved that $\breve G_{\smQ_o i}' =  G_{\smQ_o i}$ and $\breve G_{\smQ_o\smQ_o}' = 0$. Therefore the target module is invariant with $\breve G_{\smQ_o i}'' =  G_{\smQ_o i}$. \hfill $\Box$\\

%Note that the condition is satisfied for all outputs in $w_s$, not only $w_j$, thus implying that the transfers to all out-neighbors of $w_k$ remain invariant.\\
The following is the proof for Theorem 1.\\
{\bf Proof:} Expression \eqref{eqxx5} can be written as
$
w_{\smY} = \bar G^0
w_{\smD}
+ \bar H^0 \xi_{\smY}.
$ Using this expression in prediction error \eqref{eqx5} we get $
\varepsilon(t,\theta) := \bar H(q,\theta)^{-1}\left[\Delta\bar G(q, \theta)
w_{\smD}
+ \Delta\bar H(q, \theta) \xi_{\smY}\right] + \xi_{\smY}
$ where $\Delta \bar G(q,\theta) = \bar G^0 - \bar G(q,\theta)$ and $\Delta \bar H(q,\theta) = \bar H^0 - \bar H(q,\theta)$. The proof for consistency involves two steps.
\begin{enumerate}
	\item To show that $\E \varepsilon^T(t,\theta)W\varepsilon(t,\theta)$ achieves its minimum for $\Delta \bar G(\theta) = 0$ and $\Delta \bar H(\theta) = 0$,
	\item To show the conditions under which the minimum is unique.
\end{enumerate}
\textit{Step 1:} On the basis of the data generating network representation, we can write $\bar w = T^0(q)\bar\xi$ where $\bar w = [\begin{matrix} w_{\smQ}^\top & w_o^\top & w_{\smB}^\top & w_{\smA}^\top \end{matrix}]^\top$, $\bar\xi = [\begin{matrix} \bar\xi_{\smQ}^\top & \bar\xi_o^\top & \bar\xi_{\smB}^\top & \bar\xi_{\smA}^\top \end{matrix}]^\top$
%$
%\
%\underbrace{\begin{bmatrix} w_s^\top & w_o^\top & w_b^\top & w_i^\top \end{bmatrix}^\top}_{\bar w} = T^0(q) \red{\underbrace{\begin{bmatrix} \bar\xi_s^\top & \bar\xi_o^\top & \bar\xi_b^\top & \bar\xi_i^\top \end{bmatrix}^\top}_{\bar\xi}}
%%\ \ \mbox{and so also }\
%%\begin{bmatrix} w_s \\ w_b \\ w_i \end{bmatrix} = T_{sbi}^0(q) \begin{bmatrix} \xi_s \\ %\xi_o \\ \xi_b \\ \xi_i \end{bmatrix}
%$
and denote $T_{\smQ\smB\smA}^0$ as the matrix composed of the first, third and fourth (block) row of $T^0$. Substituting $T_{\smQ\smB\smA}^0$, we get $
\varepsilon(t,\theta) := \bar H(q,\theta)^{-1}\left[\Delta\bar G(q, \theta)T_{\smQ\smB\smA}^0
+ \begin{bmatrix} \Delta \bar H(\theta) & 0 & 0 \end{bmatrix}\right]\bar\xi + \varepsilon_{\smY}
$.
Let
\beq\label{eq421}
\begin{split}
	\Delta X(\theta)\bar\xi =
	\left[ \Delta \bar G(\theta) T_{\smQ\smB\smA}^0(q) + \begin{bmatrix} \Delta \bar H(\theta) & 0 & 0 \end{bmatrix} \right]\bar\xi \\
\end{split}
\eeq
The first row of the above equation is written as,
\beq \label{eq431}
\resizebox{1.0\hsize}{!}{$
\begin{split}
	&\left(\Delta G_{\smQ\smQ}(\theta)T_{\smQ\smQ}^0 + \Delta G_{\smQ\smB}(\theta)T_{\smB\smQ}^0 + \Delta G_{\smQ\smA}(\theta)T_{\smA\smQ}^0 + \Delta \tilde H_{11}(\theta)\right)\xi_{\smQ} + \\
	&\ + \left(\Delta G_{\smQ\smQ}(\theta)T_{\smQ o}^0 + \Delta G_{\smQ\smB}(\theta)T_{\smB o}^0 + \Delta G_{\smQ\smA}(\theta)T_{\smA o}^0 + \Delta \tilde H_{12}(\theta) \right)\xi_o\\
	&\ +\left(\Delta G_{\smQ\smQ}(\theta)T_{\smQ\smB}^0 + \Delta G_{\smQ\smB}(\theta)T_{\smB\smB}^0 + \Delta G_{\smQ\smA}(\theta)T_{\smA\smB}^0\right)\xi_{\smB} \\
	&\ + \left(\Delta G_{\smQ\smQ}(\theta)T_{\smQ\smA}^0 + \Delta G_{\smQ\smB}(\theta)T_{\smB\smA}^0 + \Delta G_{\smQ\smA}(\theta)T_{\smA\smA}^0\right)\xi_{\smA}.
\end{split}$
}
\eeq $\Delta \bar H(\theta)$ has a delay in each of the transfers in the matrix since both $\bar{H}(\theta)$ and $\bar H^0$ are monic. Therefore, $\Delta \tilde H_{11}(\theta)$ and $\Delta \tilde H_{12}(\theta)$ will have at least a delay in each of its transfers. By condition 5, if all $G$-elements are strictly proper and parameterized as strictly proper transfer functions, the terms in between the brackets in \eqref{eq431} has at least a delay, so that the expression \eqref{eq431} will be uncorrelated to the innovation $\xi_{\smQ}$. Otherwise, by condition 5 if the delay in path/loop condition is satisfied, the terms in $\Delta G_{\smQ\smQ}(\theta)T_{\smQ\smQ}^0, \Delta G_{\smQ\smA}(\theta)T_{\smA\smQ}^0$, $\Delta G_{\smQ\smQ}(\theta)T_{\smQ o}^0$, $\Delta G_{\smQ\smA}(\theta)T_{\smA o}^0$, $\Delta G_{\smQ\smQ}(\theta)T_{\smQ\smB}^0$, $\Delta G_{\smQ\smB}(\theta)T_{\smB\smB}^0$, $\Delta G_{\smQ\smA}(\theta)T_{\smA\smB}^0$ will have at least a delay. Also $\xi_{\smA}$ is uncorrelated to $\xi_{\smQ}$.
Therefore the expression \eqref{eq431} will be uncorrelated to the innovation $\xi_{\smQ}$.
For the second row of equation \eqref{eq421} a complete analogous situation occurs.
Therefore, when condition 1 and condition 5 are satisfied, the term $\Delta X(\theta)\bar\xi$ is uncorrelated to the innovation $\xi_{\smY}$. As a result the minimum value of $\E \varepsilon^T(t,\theta)W\varepsilon(t,\theta)$, which is $\mathbb{E}\left[\xi_{\smY}^\top W\xi_{\smY}\right]$, is achieved for $\Delta \bar G(\theta) = 0$ and $\Delta \bar H(\theta) = 0$. \\
\textit{Step 2:} When minimum is achieved we should have the power of $\varepsilon(t,\theta) - \xi_{\smY}$ to be zero, where
$
\varepsilon(t,\theta) - \xi_{\smY} = \bar H(q,\theta)^{-1}\left[\begin{bmatrix}
\Delta\bar G(q, \theta) & \Delta\bar H(q, \theta)
\end{bmatrix} \begin{bmatrix}
w_{\smD}^\top & \xi_{\smY}^\top
\end{bmatrix}^\top\right].
$
%One might have expected, based on classical results, that this spectrum condition can be rewritten in a condition on $(w_s,w_i,w_b,w_o)$ only, but this does not seem to be true in the general case. We need the $4$-dimensional signal vector in order to guarantee the above mentioned implication. We might be able though, to reformulate the term $\xi_o$ into a term $w_o$.
Using the expression of $\xi_o$ from \eqref{eq5} and substituting it in the expression of $\varepsilon(t,\theta) - \xi_{\smY}$ we get,
$
\varepsilon(t,\theta) - \xi_{\smY} = \bar H(q,\theta)^{-1}\left[\begin{bmatrix}
\Delta\bar G(q, \theta) & \Delta\bar H(q, \theta)
\end{bmatrix}J\kappa(t)\right]$ where,
\[
J = \begin{bmatrix}
I & 0 & 0 \\ 0 & I & 0 \\ -(\bar H^0_{oo})^{-1}\bar G^0_o & -(\bar H^0_{oo})^{-1}\bar H^0_{o\smQ} & 1
\end{bmatrix} \ ; \ \bar G_o^{0\top}  = \begin{bmatrix}
\bar G^{0\top}_{o\smQ} \\ \bar G^{0\top}_{o\smB} \\ \bar G^{0\top}_{o\smA}
\end{bmatrix}
\]
The standard reasoning for showing uniqueness of the identification result is to show that if the power of $\varepsilon(t,\theta) - \xi_{\smY}$ equals $0$, this should imply that $\Delta \bar G = 0$ and $\Delta \bar H = 0$. Since $J$ is full rank, writing the power of the above term in the frequency domain, through Parseval theorem, this implication will be fulfilled if
$\Phi_{\kappa}(\omega) > 0 $ for a sufficiently high number of frequencies. Thus if condition 4 is satisfied along with the other conditions in Theorem 1, it ensures that the minimum value is achieved only for $\bar G(\theta) = \bar G^0$ and $\bar H(\theta) = \bar H^0$. \hfill $\Box$

%%%%%%%%%%%%%%%%%%%%%%%%%%%%%%%%%%%%%%%%%%%%%%%%%%%%%%%%%%%%%%%%%%%%%%%%%%%%%%%%%%%%%%%%%%%%%%%
%\vspace{-0.12cm}
\bibliographystyle{IEEEtran}
\bibliography{Paul_Dynamic_Networks_Library}

\end{document}